\documentclass[aps,prd,preprint,groupedaddress,nofootinbib,longbibliography]{revtex4-1}\usepackage[T1]{fontenc}
\usepackage[latin9]{inputenc}
\usepackage{geometry}
\usepackage{amsmath}
\usepackage{amsfonts}
\usepackage{amssymb}
\usepackage{esint}
\usepackage{braket,mathtools}
\usepackage{mathrsfs}
\usepackage{stmaryrd}

\usepackage{hyperref}
\usepackage{color}

\DeclareMathAlphabet{\mathpzc}{OT1}{pzc}{m}{it}

\newcommand{\call}{{\cal L}}
\newcommand{\calo}{{\cal O}}
\newcommand{\beq}{\begin{equation}}
\newcommand{\eeq}{\end{equation}}
\newcommand{\hf}{\frac{1}{2}}
\newcommand{\ehat}{{\hat e}}
\newcommand{\bea}{\begin{eqnarray}}
\newcommand{\eea}{\end{eqnarray}}

\makeatletter
\numberwithin{equation}{section}

\makeatother

\begin{document}

\begin{titlepage}

\title{Gauge-invariant  observables, gravitational dressings, and holography in AdS}

\author{Steven B. Giddings}
\email[]{giddings@ucsb.edu}
\author{Alex Kinsella}
\affiliation{Department of Physics, University of California, Santa Barbara, CA 93106}

\date{\today}

\begin{abstract}
This paper explores construction of gauge (diffeomorphism)-invariant observables in anti de Sitter (AdS) space and the related question of how to find a ``holographic map'' providing a quantum equivalence to a boundary theory.  Observables are constructed perturbatively to leading order in the gravitational coupling by gravitationally dressing local field theory operators in order to solve the gravitational constraints. Many such dressings are allowed and two are explicitly examined, corresponding to a gravitational line and to a Coulomb field; these also reveal an apparent role for more general boundary conditions than considered previously.  The observables obey a nonlocal algebra, and we derive explicit expressions for the boundary generators of 
the SO(D-1,2) AdS isometries that act on them.  We examine arguments that gravity {\it explains} holography through the role of such a boundary Hamiltonian. Our leading-order gravitational construction reveals some questions regarding how these arguments work, and indeed construction of such a holographic map appears to require solution of the  non-perturbative generalization of the bulk constraint equations. 
\end{abstract}

\maketitle

\tableofcontents

\end{titlepage}

\section{Introduction and overview}

Two longstanding problems for gravity are how to think about gauge symmetries and about gauge-invariant observables in the quantum context.  An 
important arena for testing our understanding of quantum gravity is anti de Sitter (AdS) space.  An additional motivation for this is the widespread belief in the conjecture\cite{Mald} that quantum gravity in AdS is quantum-mechanically equivalent to a conformal field theory on its boundary.

With these motivations, this paper will extend recent work
on construction of gauge-invariant observables from the case of flat backgrounds\cite{DoGi1} to the AdS context. The basic idea is that while local operators of a field theory coupled to gravity are not gauge invariant, since diffeomorphisms relocate points, these operators may be ``gravitationally dressed'' to construct operators that are invariant under the diffeomorphism gauge symmetries.  In the perturbative theory these are the diffeomorphisms that vanish sufficiently rapidly at infinity and they  are generated by the constraints of general relativity (GR).  Colloquially, in gravity a particle is inseparable from its gravitational field, and so an operator creating a particle must also create the corresponding field.

Understanding such constructions is also interesting and important from the viewpoint of the AdS/CFT conjecture.  This is because in order for the conjecture to be true, one needs to understand how a $D$-dimensional bulk quantum-gravitational theory can be equivalent to a $D-1$-dimensional theory, through a ``holographic equivalence.''  While this was originally motivated within string theory, it has been argued by Marolf\cite{MaroUH,Marothought,Maro2013} (see also \cite{Jaco}) that the gauge structure of gravity provides the basic explanation for such a holographic correspondence, and that stringy degrees of freedom play no significant role in explaining holography.  It is clearly important to put these statements to concrete tests.
	
While we may not yet fully understand how to describe the degrees of freedom, dynamics, and symmetries of the complete theory of quantum gravity, this paper will take an approach that rests on the idea that there is a ``correspondence principle'' for quantum gravity\cite{SG2006,SGalg}:  in the long-distance/low-energy limit, for questions where gravitational fields are effectively weak, phenomena should be approximately well described by local quantum field theory coupled to perturbative general relativity.  If so, then whatever is the structure of observables in the full theory, those should match onto and be well approximated by gauge-invariant observables in the perturbative regime.  Such a ``weak-gravity correspondence''  provides an important set of constraints on the more basic theory, extending also beyond the construction of observables.  Taking this weak-field limit, we may investigate  observables and other quantities perturbatively in the gravitational coupling.  Since gauge transformations are generated by the constraints in this  limit, constructing observables involves perturbatively solving the GR constraints.  This paper takes such a perturbative approach.

In outline, the next section will discuss the explicit construction of two different kinds of gauge-invariant observables to leading order in the gravitational coupling.  The first is a gravitational line dressing; the dressed operator is shown to create a particle together with a gravitational field concentrated to a narrow region extending to infinity.   The second is a Coulomb dressing, which creates a symmetric gravitational field; this matches the AdS-Schwarzschild solution, although with non-standard boundary conditions on asymptotic field behavior.  The difference between these field configurations is a radiation field; the state created by the line dressing decays to the Coulomb field plus radiation at infinity.  These are examples of a very large set of possible gravitational dressings.\footnote{For related constructions in AdS${}_3$, see \cite{DMM,Guic,ACFLKL,CFKL}.}  In general, the presence of the gravitational dressing means that  observables obey a nonlocal algebra, as in flat space \cite{SGalg,DoGi1}, whose nonlocality can be characterized by the locality bound of \cite{GiLi1,GiLi2,SGloc}.

Acting on such operators that commute with the constraints, generators of the global $SO(D-1,2)$ symmetries of AdS reduce to surface terms.  Using the covariant canonical formalism described in \cite{ashtekar1982,Crnkovic:1986ex,Ashtekar:1987hia,Zuckerman:1989cx,Crnkovic:1987tz,Lee:1990nz,Iyer:1994ys,DoGi2}, section III derives an explicit expression for these surface charges.  The Hamiltonian and rotation generators are in particular examined in detail and related to components of a ``boundary stress tensor'' defined in terms of the asymptotic  metric perturbations.   We also explicitly check that this boundary Hamiltonian correctly generates time translation of the gauge-invariant operators.

Section IV turns to the question of a gravitational explanation for holography.  The presence and role of the gravitational dressings raises interesting puzzles about how gravity might explain holography.  First, the ambiguity in dressings for gauge-invariant operators appears to lead to an ambiguity in the ``extrapolate'' dictionary, where a bulk operator is supposed to correspond to a boundary operator which is taken to the boundary and rescaled.  This is part of the complication in trying to construct an inverse mapping that determines the general bulk operator in terms of boundary operators, as in \cite{BDHM,HKLL,HMPS,KLgauge,KLgrav}.   But, beyond that, we explicitly see that 
the ``boundary unitarity''  and ``holography of information'' arguments of \cite{MaroUH,Maro2013}, and particularly construction of an equal time ``holographic map'' determining bulk operators in terms of boundary operators, appear to rely on non-perturbatively solving the constraints, or their non-perturbative generalization, which involves solving for  unitary bulk evolution.  In this sense, an AdS/CFT correspondence doesn't solve the problem of unitary bulk evolution, but rather existence of a holographic map appears to require its solution.

Appendices briefly outline canonical quantization of perturbative gravity in AdS, extension of the dressing theorem of \cite{DoGi2} to AdS, and discussion of boundary conditions on gravitational perturbations.

\section{\label{DIO}Diffeomorphism-invariant observables in AdS}

\subsection{Basic setup}

One of the goals of this paper is to explore diffeomorphism-invariant (gauge-invariant) operators that create excitations in the bulk of AdS.  In doing this, we will explore the implications of gauge invariance in a perturbative expansion in $G_D$, the gravitational coupling.  Specifically, to help understand the basic structural issues, we will consider perturbative quantization of a theory with a scalar field minimally coupled to gravity in AdS, and so work with the action
\beq\label{gravact}
S= \int d^D x \sqrt{|g|}\left\{ \frac{2}{\kappa^2} \left(R -2 \Lambda \right)+\call_{gf} - \hf \left[(\nabla \phi)^2 + m^2 \phi^2 \right]\right\}\ ,
\eeq
where $\kappa^2=32\pi G_D$ and $\Lambda$ is the cosmological constant.  Here $\call_{gf}$ is a possible ``gauge-fixing'' (really, gauge-breaking) term; a common choice, working in a background $g^0$, is
\beq\label{Lgf}
\sqrt{|g|} \call_{gf}=-\frac{1}{\alpha\kappa^2}\frac{\sqrt{|g^0|}}{|g|}\left[\nabla^0_\mu\left(\sqrt{|g|} g^{\mu\nu}\right)\right]^2\ ,
\eeq
with $\alpha$ the gauge-breaking parameter.  Indices $\mu$, $\nu$, {\it etc.} run over the $D$ bulk spacetime dimensions.  This paper will consider the AdS background metric
\begin{equation}\label{adsmet}
ds_0^{2}=\frac{R^{2}}{\cos^{2}\rho}\left(-d\tau^{2}+d\rho^{2}+\sin^{2}\rho\ d\Omega_{D-2}^{2}\right)
\end{equation}
where $R=\sqrt{-(D-1)(D-2)/2\Lambda}$ is the AdS radius and $d\Omega_{D-2}^{2}$ is the round metric on the $(D-2)$-sphere. 

We will use the approach of \cite{DoGi1} to construct gauge-invariant observables in AdS, working in the perturbative expansion
\beq\label{metpert}
g_{\mu\nu} = g^0_{\mu\nu} + \kappa h_{\mu\nu}\ .
\eeq  
At linear order in $\kappa$, a diffeomorphism with parameter $\kappa \xi^\mu$ transforms the fields by
\bea\label{gaugexm}
\nonumber \delta_{\kappa\xi} \phi(x) &&= -\kappa \call_\xi \phi + \calo(\kappa^2) = -\kappa \xi^\mu \partial_\mu \phi + \calo(\kappa^2)\quad ,\\ 
\quad \delta_{\kappa\xi} h_{\mu\nu} &&= -\call_\xi g_{\mu\nu} + \calo(\kappa)= -\nabla_\mu\xi_\nu -\nabla_\nu\xi_\mu + \calo(\kappa)\ 
\eea
where the covariant derivative is that of the background $g^0$.
So, at this order the field operator $\phi(x)$ is not gauge-invariant.  However, a gauge-invariant modification to this operator can be found, by ``dressing'' it with its gravitational field.

Gauge invariance of such operators may be checked either by examining their commutators with the gravitational constraints, or by examining their explicit gauge dependence.  There are in fact many different such gauge-invariant dressings of a given operator $\phi(x)$; roughly, these 
dressings differ by operators creating gravitational fields that are source-free solutions of the gravitational equations.  At linear order, for example, different dressings include a Coulomb-like dressing and a line-like dressing, both described in flat space in \cite{DoGi1}.  The resulting gauge-invariant operators can be written in the form
\beq\label{giop}
\Phi(x) = \phi(x^\mu + V^\mu(x))\ ,
\eeq
where $V^\mu(x)$ is a functional of the metric such that $\Phi(x)$ is gauge-invariant.

\subsection{Gravitational line dressing}

\subsubsection{Construction}

Here, as a first example, we will construct  line-like dressings analogous to those of \cite{DoGi1}.  Like there, we can think of doing so geometrically.  We take the boundary of AdS to serve as a fixed ``platform,'' where the diffeomorphisms vanish.  If we seek to define a gauge-invariant operator
$\Phi_L$, in a general perturbation \eqref{metpert} of AdS, we can do so by locating a field operator $\phi$ at a point we find by 
by picking a point $(\tau, \ehat)$ on the boundary ($\ehat$ is a unit vector giving a point on $S^{D-2}$) and launching a geodesic perpendicular to the boundary for a given distance.  The coordinates $(\tau, \ehat)$ and this distance specify the gauge-invariant location of $\phi$.  If we use these parameters as coordinates, the metric perturbation satisfies
\beq
h_{\rho\mu}=0\ ,
\eeq
which we can think of as a specification of an axial, gaussian-normal, or in the AdS context, Fefferman-Graham (FG), gauge.  Then to write $\Phi_L(x)$ in terms of arbitrary coordinates, following \cite{DoGi1}, we need to undo the gauge transformation taking us to this FG gauge.  Given the gauge transformation \eqref{gaugexm}, FG gauge is found from a general perturbed metric by solving the equation
\begin{equation}\label{FGxm}
\nabla_{\mu}\chi_{\rho}+\nabla_{\rho}\chi_{\mu}=h_{\rho\mu}\ .
\end{equation}
So, given this solution, the gauge-invariant operator $\Phi_L(x)$ will take the form \eqref{giop} with 
\beq\label{VL}
V^\mu_L=-\kappa \chi^\mu\ .
\eeq

\subsubsection{Explicit solution}

Solving the equation \eqref{FGxm} thus gives the needed dressing, and is straightforward.  First we rewrite \eqref{FGxm} as a collection of ordinary differential equations,
 \begin{equation}
\partial_{\mu}\chi_{\rho}+\partial_{\rho}\chi_{\mu}-2\Gamma_{\rho\mu}^{\lambda}\chi_{\lambda}=h_{\rho\mu}\ .
\end{equation}
  The nonvanishing Christoffel symbols for the metric  \eqref{adsmet} are given by
\beq
\Gamma_{\tau\tau}^{\rho}=\Gamma_{\rho\tau}^{\tau} =\Gamma_{\rho\rho}^{\rho}= \tan\rho\quad ,\quad 
\Gamma_{ab}^{\rho} = - \tan\rho\,\hat g_{ab}\quad ,\quad
\Gamma_{\rho b}^{a} = \frac{1}{\sin\rho\cos\rho}\delta^a_b\quad ,\quad \Gamma^a_{bc}=\hat \Gamma^a_{bc} \ 
\eeq
where $a,b,\cdots$ are indices\footnote{We collect our index conventions here:  $\mu,\nu,\cdots$ denote bulk $D$-dimensional spacetime indices; $i,j,\cdots$ denote bulk $D-1$-dimensional spatial indices; $\alpha,\beta,\cdots$ denote boundary $D-1$-dimensional spacetime indices; and $a,b,\cdots$ denote boundary $D-2$-dimensional spatial indices.} and $\hat g_{ab}$ is the unit-radius metric on the boundary $S^{D-2}$.
The $\rho$ equation has a similar structure to the flat space case and is immediately integrated to find $\chi^\rho$, and thus via \eqref{VL}, 
\beq\label{rcomp}
V_L^{\rho}(x)=\kappa \frac{\cos\rho}{2R^2}\int_{\rho}^{\pi/2}du\cos u\ h_{\rho\rho}(\tau,u,\ehat)
\eeq
where the coordinates  of the point $x^\mu = (\tau, \rho, \ehat)$ determine the unit spatial vector $\ehat$; that is, integration is along a radial line from $x$ to the boundary.  

Next, the $\tau$ equation takes the form 
\begin{equation}
\partial_{\rho}\chi_{\tau}-2\tan\rho\chi_{\tau}=h_{\tau\rho}-\partial_{\tau}\chi_{\rho}\ ;
\end{equation}
this can be integrated to give, via \eqref{VL}, 
\beq\label{tcomp}
V_L^{\tau}(x)=-\frac{\kappa}{R^2}\left[\int_{\rho}^{\pi/2}du\left( \cos^{2}u\,h_{\rho\tau}(\tau,u,\ehat)
+\frac{\cos u}{2}\int_{u}^{\pi/2}du'\cos u'\ \partial_{\tau}h_{\rho\rho}(\tau,u',\ehat ) \right)\right]\ .
\eeq
Finally, equations for the angular components take the form
\begin{equation}
\partial_{\rho}\chi_{a}-\frac{2}{\sin\rho\cos\rho}\chi_{a}=h_{a\rho}-\partial_{a}\chi_{\rho}\ ,
\end{equation}
and integrate to give, via \eqref{VL}, 
\beq\label{acomp}
V_L^{a}(x) = \frac{\kappa}{R^2}{\hat g}^{ab} \left[\int_{\rho}^{\pi/2}du\ \cot^{2}u\left(h_{\rho b}(\tau,u,\ehat)+\frac{1}{2\cos u}\partial_{b}\int_{u}^{\pi/2}du'\ h_{\rho\rho}(\tau, u',\ehat )\cos u'\right)\right]\ .
\eeq
The expressions \eqref{rcomp}, \eqref{tcomp}, and \eqref{acomp} thus specify the line dressing version of the general dressed operator 
\eqref{giop}.

\subsubsection{Gauge invariance}

Gauge invariance of $\Phi(x)$ of the resulting expression \eqref{giop} is readily checked.  Given equations \eqref{rcomp}, \eqref{tcomp}, and \eqref{acomp}, and the gauge transformation \eqref{gaugexm}, one straightforwardly finds the gauge transformation
\beq\label{Vtrans}
\delta_{\kappa \xi} V_L^\mu(x) = \kappa \xi^\mu\ .
\eeq
This, to linear order, exactly cancels the gauge transformation of $\phi$ from \eqref{gaugexm}, in the expression for $\Phi(x)$.  

This can alternately be phrased in terms of the generators of gauge transformations, which are  diffeomorphisms vanishing at infinity and are generated by  the constraints, $G_0^\mu+\Lambda \delta^\mu_0 - 8\pi G T_0^\mu$ (see the next section).  The linearized versions of these should commute with the operator $\Phi(x)$.  Alternately, in the presence of a gauge-breaking term \eqref{Lgf}, these gauge transformations are generated by the linearized version of the gauge-fixing term $\nabla^0_\mu\left(\sqrt{|g|} g^{\mu\nu}\right)$.

\subsubsection{Dressing field}\label{Dressf}

It is also instructive to find the gravitational field created by a given dressing operator; this was called the ``dressing field'' in \cite{DoGi1}. This is found by computing a commutator
\beq \label{dresscom}
[h_{\mu\nu}(x),\Phi(x')]\ ;
\eeq
$\Phi$ also in general yields a non-zero value for $\dot h_{\mu\nu}$, found via the commutator with the conjugate momentum.  For simplicity we consider the case of a very massive particle, at the center of AdS (in some frame), so that the momentum may be neglected.  The commutator \eqref{dresscom} then becomes (from the creation part of the operator)
\beq
[h_{\mu\nu}(x),\Phi(0)] \simeq imR [h_{\mu\nu}(x),V_L^\tau(0)]\,\phi(0)\ ,
\eeq
so in this limit the dressing field is given by the expression
\beq
{\tilde h}_{\mu\nu}(x) \simeq imR [h_{\mu\nu}(x),V_L^\tau(0)]\ .
\eeq

This field can be evaluated from the dressing using commutation relations for the metric, which are worked out in  Appendix \ref{CQAdS}.
Gauge invariance of $\Phi$ indicates that the result should be independent of gauge parameter; we set $\alpha=\infty$, corresponding to the symmetry-restoration limit.  The expression for the commutator simplifies if we first integrate the second term of \eqref{tcomp} by parts, to find
\beq\label{newVL}
V_L^\tau(0)= -\frac{\kappa}{R^2} \int_0^{\pi/2} du \left( \cos^2 u\, h_{\rho\tau} + \frac{\sin u \cos u}{2} \dot h_{\rho\rho}\right)\ .
\eeq 
The resulting non-zero components of the dressing field are, at $\tau=0$, 
\beq\label{linedress}
{\tilde h}_{\rho\rho}(0,\vec x) \simeq \frac{D-3}{2(D-2)}\frac{\kappa m}{R^{D-5}} \frac{\cos^{D-5}\rho}{\sin^{D-3}\rho} \frac{\delta(\hat e -\hat e')}{\sqrt{\hat g}}\ 
\eeq
and
\beq\label{linedressa}
{\tilde h}_{ab}(0,\vec x) \simeq -\frac{1}{2(D-2)}\frac{\kappa m}{R^{D-5}} \frac{\cos^{D-5}\rho}{\sin^{D-5}\rho} \frac{\delta(\hat e -\hat e')}{\sqrt{\hat g}}{\hat g}_{ab}\ 
\eeq
where $\hat e'$ points along the gravitational line.

For the line dressing these are concentrated on  an infinitesimally thin line; this behavior may be regulated by averaging over a small solid angle $\Delta \Omega$.  To do that, we replace \eqref{newVL} by 
\beq\label{lineav}
V_\Delta^\tau(0)=-\frac{\kappa}{R^2 |\Delta \Omega_{D-2}|}\int_0^{\pi/2} du \int_{\Delta \Omega_{D-2}}d\Omega_{D-2} \left( \cos^2 u\, h_{\rho\tau} + \frac{\sin u \cos u}{2} \dot h_{\rho\rho}\right)\ , 
\eeq
resulting in a dressing field smeared over the solid angle $\Delta \Omega_{D-2}$.

While the gravitational line dressing given in this section gives a consistent description of gauge-invariant operators, to linear order in $\kappa$, the dressing field that it produces is not a static field.  Specifically, the dressed operator $\Phi_L(x)$ creates a particle together with a gravitational field localized to the line; in subsequent evolution the non-trivial gravitational field will spread out\cite{DoGi1}, analogous to the behavior found in electrodynamics\cite{Shab,PFS,HaJo}.   One expects that after radiating gravitational radiation to infinity, the field  ultimately settles down to a more symmetric Coulomb field.  We next turn to the operator that creates this field directly.

\subsection{Coulomb dressing}

\subsubsection{Construction}

As in \cite{DoGi1}, we expect to also be able to construct a Coulomb-like dressing by averaging the gravitational line dressing over all angles.  
The $\tau$ component of the dressing is given by \eqref{lineav}, integrated over the full solid angle $\Omega_{D-2}$ to give
\beq\label{Coult}
V_C^\tau(0)=-\frac{\kappa}{R^2 \Omega_{D-2}}\int du d\Omega_{D-2} \left( \cos^2 u\, h_{\rho\tau} + \frac{\sin u \cos u}{2} \dot h_{\rho\rho}\right)\ .
\eeq
The spatial part of the Coulomb dressing is most easily determined by starting with an expression analogous to that in \cite{DoGi1},
\beq
V^i_C(0) = \int dV_{D-1} f(\rho) E^i E^\mu E^\nu h_{\mu\nu}\ .
\eeq
Here $dV_{D-1}$ is the spatial volume element for the AdS metric \eqref{adsmet}, $E^\mu=(0, \cos\rho/R,0\cdots)$ is the unit radial vector, and $f(\rho)$ is chosen so that the dressing also transforms as needed, \eqref{Vtrans}, under a diffeomorphism \eqref{gaugexm}; correspondingly it has normalization matching the flat space result\cite{DoGi1}.  These determine $f(\rho)=\kappa (D-1)/(2 R^{D-2} \tan^{D-2}\rho\, \Omega_{D-2})$, so that
\beq
V^i_C(0) = \frac{\kappa (D-1)}{2\Omega_{D-2}} \int  \frac{Rd\rho}{\cos\rho} d\Omega_{D-2}  E^i E^j E^k h_{jk}\ .
\eeq

\subsubsection{Coulomb dressing field and relation to AdS-Schwarzschild}

The Coulomb dressing field is computed in analogy with that of the line dressing, above.  Indeed, the non-trivial components of the metric at $\tau=0$ may be found directly from the angle-averaged formula \eqref{Coult}, which yields (see \eqref{linedress})
\beq\label{dressfieldr}
{\tilde h}_{\rho\rho}(x) \simeq imR [h_{\rho\rho}(x),V_C^\tau(0)] 
= \frac{D-3}{2(D-2)\Omega_{D-2}} \frac{\kappa m}{R^{D-5}} \frac{\cos^{D-5}\rho}{\sin^{D-3}\rho}\ ,
\eeq
One likewise finds
\beq\label{dressfielda}
{\tilde h}_{ab}(x) \simeq imR [h_{ab}(x),V_C^\tau(0)]= -\frac{1}{2(D-2)\Omega_{D-2}} \frac{\kappa m}{R^{D-5}} \cot^{D-5}\rho {\hat g}_{ab}\ .
\eeq

These perturbations may be compared with the expected Schwarzschild mass perturbation of AdS.  This is found from the standard expression
\beq
ds^2= - \left(1+\frac{r^2}{R^2} - \frac{K_D m}{r^{D-3}}\right)dt^2 + \left(1+\frac{r^2}{R^2} - \frac{K_D m}{r^{D-3}}\right)^{-1} dr^2 + r^2 d\Omega_{D-2}^2\ 
\eeq
with constant
\beq
K_D=\frac{\kappa^2}{2(D-2)\Omega_{D-2}}\ .
\eeq
Using the transformation $t= R\tau$, $r=R \tan\rho$, and expanding to linear order in $\kappa^2$, this yields the Schwarzschild metric perturbation
\beq\label{Schpert}
\kappa h_{\mu\nu}dx^\mu dx^\nu= \frac{K_D m}{R^{D-5}} \cot^{D-3}\rho (d\tau^2 + d\rho^2)\ .
\eeq

There is clear disagreement between the dressing field \eqref{dressfieldr}, \eqref{dressfielda} and the Schwarzschild perturbation \eqref{Schpert}; in fact the former don't even have the na\"\i vely expected\cite{HeTe} $\cos^{D-3}\rho$ falloff behavior as $\rho\rightarrow\pi/2$.\footnote{Brief discussion of  falloff behavior and finiteness and conservation of the symplectic form is given in Appendix \ref{Falloffs}.}  However, the linear perturbations \eqref{dressfieldr}, \eqref{dressfielda} are related to \eqref{Schpert} through a diffeomorphism 
\beq
\tilde\rho= \rho+ \frac{K_D m}{2 R^{D-3} } \frac{\cos^{D-2}\rho}{\sin^{D-4}\rho} \ ,
\eeq
establishing that the dressing does indeed create a field of the correct form.

\subsection{Dressing ambiguity and nonlocal algebra}

The preceding subsections have illustrated two different dressings for the field operator $\phi(x)$.  The dressing fields that these create differ by a solutions of the homogeneous (source-free) gravitational equations, that is, by a radiation field.  Specifically, as we have noted, the line dressing field \eqref{linedress}, \eqref{linedressa} will evolve into a Coulomb field plus radiation to infinity\cite{DoGi1}, in parallel with the QED case\cite{Shab,PFS,HaJo}.  More generally, we might expect there to be a very large number of gauge-invariant dressings, corresponding to all the possible radiation fields by which the Coulomb field can be augmented.  For example, \cite{LTV} suggest dressings that are $Z_2$ symmetric, which may represent yet another prescription within this wide ambiguity; it would be interesting to further test their proposal via comparison with the type of construction we have outlined.  Note also that we expect to be able to extend such dressing fields to consistent solutions at higher orders in $\kappa$.  In particular, work by Carlotto and Schoen\cite{CaSc} (for a review, see \cite{Chru}) shows that there are initial data for the full nonlinear Einstein equations which vanish outside of specified cones, suggesting a way to extend dressing fields similar to \eqref{lineav} beyond the linear approximation, and to generalize to a distribution of conical fields.

The gauge invariance of gravity and the need to solve the constraints implies nonlocality of the observables and of their algebra in gravity.  A characterization of when this becomes important is the ``locality bound'' of \cite{GiLi1,GiLi2,SGloc}.  Non-commutativity of observables due to dressing in flat space was explicitly shown in \cite{DoGi1}, confirming the locality bound.\footnote{Prior to that, non-commutativity in a gauge-fixed approach was studied in \cite{Heem}.  Refs.~\cite{KLgauge,KLgrav} also discussed the algebra of observables in gauge theory and gravity in AdS, but did not exhibit the non-commutativity we describe.}  

The expressions for the dressing that we have derived for AdS likewise exhibit the nonlocality of the observables and their algebra in this context.  Specifically, while the commutator of the scalar field $\phi$ vanishes at spacelike separation, commutators of gauge-invariant observables like \eqref{giop} do not in general vanish at spacelike separation.  This clearly occurs either for the line form or Coulomb form of the dressing $V^\mu(x)$, and nonzero commutators analogous to the expressions in \cite{DoGi1} can be worked out.

Thus, as in flat space, the observables of a gravitational theory in AdS do not obey a local algebra.  This
has potentially important implications for the nature of locality in quantum gravity, and in particular it obstructs an algebraic definition of locality analogous to that in quantum field theory\cite{Haag}; an alternate approach to characterizing localization has been preliminarily discussed in \cite{DoGi3}.  

It has also been argued that this gravitational nonlocality is at the heart of holography, and explains its existence\cite{MaroUH,Marothought,Maro2013} (see also \cite{Jaco}).  Specifically, the constraints tell us that the Hamiltonian is a boundary term, suggesting that evolution can be completely characterized in terms of evolution at the boundary.  A boundary Hamiltonian can generate time translation precisely because of the nonlocality of the observables that we have described.  We turn next to examining this question more closely, starting with a derivation of the generators of the $SO(D-1,2)$ symmetries of AdS.

\section{Symmetry generators and the ``boundary stress tensor''}

It was found in \cite{DoGi1} that in flat space the presence of the gravitational dressing is precisely what is needed so that generators of Poincar\'e transformations, which are surface terms in general relativity, act to correctly transform the dressed fields.  Indeed, \cite{DoGi2} proved a flat space ``dressing theorem,'' stating that once a local operator is gravitationally dressed to make it gauge-invariant, that dressing must involve the asymptotic metric; this asymptotic dependence is what is needed to ensure the correct commutators with these surface terms.  In order to understand the analogous story for AdS, one needs to find the analogous generators of the $SO(D-1,2)$ isometries of AdS, which likewise will be surface terms at the boundary of AdS.  This section will investigate these generators; for completeness, the AdS version of the dressing theorem is given in Appendix B.

The generators of  $SO(D-1,2)$ and their relation to what has been called the ``boundary stress tensor'' have been discussed in the literature; we seek an explicit expression for them in terms of the metric perturbation.  This can be worked out following the canonical covariant approach developed in \cite{ashtekar1982,Crnkovic:1986ex,Ashtekar:1987hia,Zuckerman:1989cx,Crnkovic:1987tz,Lee:1990nz,Iyer:1994ys}, which is summarized in appendix B 
of \cite{DoGi2}.  

\subsection{Symmetry generators: derivation}

Consider a general diffeomorphism $\xi^\mu$.  As is reviewed in \cite{DoGi2}, this has generator $H_\xi$ found by solving the equation
\beq\label{deltaH}
\delta H_\xi = \delta \left( \int_\Sigma C_\xi + \oint_{\partial \Sigma} Q_\xi \right) -  \oint_{\partial \Sigma} i_\xi \theta\ .
\eeq
Here $\delta$ denotes a variation (exterior derivative) on field space, $\Sigma$ is a Cauchy surface, and $C_\xi$ is a $D-1$ form whose Hodge dual (in the conventions of \cite{DoGi2}, appendix A) is proportional to the Einstein equations,
\beq\label{Ceq}
(\star C_\xi)_\mu = \left(T_{\mu\nu}-\frac{4\Lambda}{\kappa^2} g_{\mu\nu} - \frac{4}{\kappa^2} G_{\mu\nu}\right) \xi^\nu \ .
\eeq
The Noether charge $Q_\xi$ $D-2$-form and symplectic potential $\theta$ are given by 
\begin{eqnarray}
Q_{\xi} & = & -\frac{2}{\kappa^2}\star d\xi\\
\label{sympot}(\star\theta)_{\mu} & = & -\frac{2}{\kappa^2}\left(\nabla^{\nu}\delta g_{\mu\nu}-\nabla_{\mu}\delta g\right)\ ,
\end{eqnarray}
 with $\xi_\mu dx^\mu=g_{\mu\nu}\xi^\nu dx^\mu$, and $i_\xi$ is the standard interior product.  Recall that the symplectic form is $\Omega = \int_\Sigma \delta \theta$; to enforce its finiteness and conservation we restrict to dimensions $D\geq4$.\footnote{For further discussion, see Appendix \ref{Falloffs}.}  If the condition 
 \beq
 \oint_{\partial \Sigma} i_\xi \delta\theta =0
 \eeq
 is satisfied (see B21 of \cite{DoGi2}), 
ref.~\cite{DoGi2} argues that
\beq
\oint_{\partial \Sigma} i_\xi \theta = \delta \left( I_{\kappa h} \oint_{\partial\Sigma} i_\xi \theta\right)\ ,
\eeq
where the field-space interior product $I_{\kappa h}$ instructs us to replace the infinitesimal variation $\delta g$ by $\kappa h$.  
Then, eq.~\eqref{deltaH} can be solved for the generator of $\xi$,
\beq\label{Hdef}
H_\xi = I_{\kappa h}\oint_{\partial \Sigma} ( \delta Q_\xi -i_\xi\theta) + \int_\Sigma C_\xi\ .
\eeq
For a solution of the gravitational constraint equations, 
\beq\label{Constraints}
n^\mu (\star C_\xi)_\mu=0\ ,
\eeq
with $n^\mu$ the unit normal to $\Sigma$, 
the last term vanishes, making the generator a surface term, as stated.  All that remains is to evaluate the remaining expressions in \eqref{Hdef}, to derive the explicit form of the resulting charges.

To evaluate the first term, we can rewrite
\beq
(Q_\xi)_{\mu_1\cdots\mu_{D-2}}= -\frac{2}{\kappa^2} \epsilon_{\mu_1\cdots\mu_{D-2}\nu\lambda} g^{\nu\sigma} g^{\lambda\omega}\partial_\sigma(g_{\omega\zeta} \xi^\zeta)\ .
\eeq
Then, varying $g$ and replacing $\delta g$ by $\kappa h$ gives
\beq
(I_{\kappa h} \delta Q_\xi)_{\mu_1\cdots\mu_{D-2}} = -\frac{2}{\kappa}  \epsilon_{\mu_1\cdots\mu_{D-2}\nu\lambda} \left(\frac{h}{2} \nabla^\nu\xi^\lambda - h^{\nu\sigma} \nabla_\sigma \xi^\lambda + \nabla ^\nu h^{\lambda\omega}\xi_\omega\right)\ .
\eeq
The second term gives
\beq
(I_{\kappa h} i_\xi \theta)_{\mu_1\cdots\mu_{D-2}} = \frac{2}{\kappa}  \epsilon_{\mu_1\cdots\mu_{D-2}\nu\lambda}\left(\nabla_\sigma h^{\nu\sigma} - \nabla^\nu h\right)\xi^\lambda\ .
\eeq
Combining these gives the generator
\bea\label{gengen}
H^\partial_\xi=\frac{2}{\kappa} \oint d^{D-2} \Omega \lim_{\rho\to \pi/2} (R\tan\rho)^{D-2}\,
&&\frac{\delta_{(\tau\rho)}^{\nu\lambda} \cos^2\rho}{R^2}\Biggl[\frac{h}{2} \nabla_\nu\xi_\lambda - h_\nu^{\sigma} \nabla_\sigma \xi_\lambda \cr
&&+ \nabla_\nu h_{\lambda\omega}\xi^\omega+\left(\nabla_\sigma h_\nu^{\sigma} - \nabla_\nu h\right)\xi_\lambda\Biggr]
\eea
where $\delta_{(\tau\rho)}^{\nu\lambda}$ is the unit antisymmetric symbol for indices $\tau,\rho$.  

\subsection{Boundary Hamiltonian}\label{bdyham}

For example, consider the case where $\xi^\mu=(1,\vec0)$ is the Killing vector generating time translations.  Then, one can straightforwardly find the Hamiltonian generator
\beq\label{Hbdyp}
H^\partial=\frac{2}{\kappa} \oint d^{D-2}\Omega  \hat n^\tau \xi^\tau  \lim_{\rho\to \pi/2}(R \tan\rho)^{D-2} \left(\nabla_a h^a_\rho - \nabla_\rho h^a_a+h^a_a \tan \rho\right)\ .
\eeq
Here the covariant derivatives are calculated with the bulk metric, and the area element and normal $\hat n^\mu=(1,\vec0)$ are defined using the boundary metric,\footnote{Index conventions are summarized in an earlier footnote.}
\beq
d\hat s^2 = {\hat g}_{\alpha\beta}dx^\alpha dx^\beta = -d\tau^2 + d\Omega^2\ .
\eeq
$H^\partial$ may also be rewritten in terms of the boundary metric and its covariant derivative ${\hat \nabla}_\alpha$:
\bea\label{Hbdy}
H^\partial=\frac{2}{\kappa} \oint d^{D-2}\Omega\,  \hat n^\tau \xi^\tau  \lim_{\rho\to \pi/2} (R\tan\rho)^{D-4} \Biggl[&&\hat \nabla^a h_{a\rho} +(D-2)\tan\rho\,h_{\rho\rho}\cr && +\left(\frac{2-\cos^2\rho}{\sin\rho\cos\rho}-\partial_\rho\right) \hat g^{ab} h_{ab}\Biggr]\ .
\eea

Since this Hamiltonian is a boundary integral, it is natural to propose that the integrand be identified with a ``boundary stress tensor,''
\beq\label{bdyT}
{\cal T}_{\tau\tau} = \frac{2}{\kappa}\lim_{\rho\to \pi/2} (R\tan\rho)^{D-4} \left[\hat \nabla^a h_{a\rho} +(D-2)\tan\rho\,h_{\rho\rho} +\left(\frac{2-\cos^2\rho}{\sin\rho\cos\rho}-\partial_\rho\right) \hat g^{ab} h_{ab}\right]\ .
\eeq 
However, note that the metric doesn't always obey the boundary conditions that have been  typically assumed for normalizable perturbations\cite{HeTe},
\beq\label{HeTeBC}
h_{\alpha\beta}\rightarrow \cos^{D-3}\rho\, \mathpzc{h}_{\alpha\beta}(x^\gamma)
\quad,\quad h_{\rho\rho}\rightarrow \cos^{D-3}\rho\, \mathpzc{h}_{\rho\rho}(x^\alpha)
 \quad,\quad h_{\rho\alpha}\rightarrow \cos^{D-2}\rho\, \mathpzc{h}_{\rho\alpha}(x^\beta)\ ,
\eeq
and so ${\cal T}_{\tau\tau}$ is not trivially reexpressed in terms of such coefficients of the asymptotic metric.  (Further discussion of normalizability and boundary 
conditions appears in Appendix \ref{Falloffs}.)
Specifically, the Coulomb fields \eqref{dressfieldr} and \eqref{dressfielda} have asymptotic behavior $h_{\rho\rho}\sim h_{ab}\sim \cos^{D-5}\rho$, so with an extra power of $1/\cos^2\rho$.  Nonetheless, one may check that when \eqref{Hbdy} is evaluated for these perturbations, it gives the correct answer, 
\beq
H^\partial=mR\ ;
\eeq
the na\"\i vely singular behavior in \eqref{Hbdy} cancels between the terms.

Note that while we have been lead to consider boundary conditions more general than \eqref{HeTeBC}, in the special case where these boundary conditions are assumed, the Hamiltonian and stress tensor can be written in terms of the metric coefficients appearing in \eqref{HeTeBC}.  In this case, the $\rho\rightarrow\pi/2$ limit gives
\beq\label{Hamil}
 H^\partial= \frac{2}{\kappa} \oint d^{D-2}\Omega  \hat n^\tau \xi^\tau\ R^{D-4} [(D-2)\mathpzc{h}_{\rho\rho} +(D-1)\hat g^{ab} \mathpzc{h}_{ab}]\ , 
 \eeq
and
\beq
{\cal T}_{\tau\tau} =\frac{2R^{D-4}}{\kappa}[(D-2)\mathpzc{h}_{\rho\rho} +(D-1)\hat g^{ab} \mathpzc{h}_{ab}]\ .
\eeq

\subsection{Rotation generators}

We can likewise derive expressions for other $SO(D-1,2)$ generators.  For example, consider the rotation generators.  For simplicity, choose coordinates so that the rotation is  in the last, azimuthal, angle, which we call $\phi$, and so gives Killing vector $\eta=\frac{\partial}{\partial\phi}$.  This can then be used in the expression \eqref{gengen}, which is found to reduce to
\bea
H^\partial_{\eta}  &&=  \frac{2}{\kappa}\oint d^{D-2}\Omega{\hat n}^\mu \eta^{\nu}\lim_{\rho\to\pi/2} R^{D-4}\tan^{D-2}\rho\ \left(2h_{\mu\nu}\cot\rho+2 \cos^2\rho\, \nabla_{[\mu}h_{\rho]\nu}\right)\cr &&= \frac{4}{\kappa}\oint d^{D-2}\Omega{\hat n}^\mu \eta^{\nu} \lim_{\rho\to\pi/2} R^{D-4}\tan^{D-2}\rho \left( \cos^2\rho\, \partial_{[\mu} h_{\rho]\nu} + \cot \rho\, h_{\mu\nu}\right)
\eea
with antisymmetrization normalization $[ij]=(ij-ji)/2$.
This leads to the proposal that the $\tau a$ component of the ``boundary stress tensor'' is
\beq\label{Ttf}
{\cal T}_{\tau a}= \frac{4}{\kappa} \lim_{\rho\to\pi/2} R^{D-4}\tan^{D-2}\rho \left( \cos^2\rho\, \partial_{[\tau} h_{\rho]a} + \cot \rho\, h_{\tau a}\right)\ .
\eeq

\subsection{Translating bulk operators}\label{TBO}

In order to better understand the bulk and boundary algebras, and their relation, we also want to check that  the boundary hamiltonian of section \ref{bdyham} acts to translate the dressed field $\Phi(x)$.  Here, as expected, the role of the dressing is critical.  We also would like to more clearly understand the form of the hamiltonian acting on expressions that are not gauge-invariant.

The full hamiltonian is given by eq.~\eqref{Hdef}, with $\xi^\mu=(1,{\vec 0})$.  Given  $C_\xi$ from \eqref{Ceq} and the boundary hamiltonian $H_\partial$ from eq.~\eqref{Hbdyp} or \eqref{Hbdy}, this takes the form
\beq
H=H^\partial + \int_\Sigma dV n^\mu \left[ T_{\mu\nu} - \frac{1}{8\pi G}\left(G_{\mu\nu} +\Lambda g_{\mu\nu}\right)  \right]\xi^\nu
\eeq
where $dV$ is the $D-1$ volume element and $n^\mu$ the unit normal to $\Sigma$.  $H^\partial$ is evaluated on $\partial\Sigma$.  The dressing makes $\Phi(x)$ gauge-invariant, so it will commute with the term in square brackets, proportional to the constraints \eqref{Constraints}.  
Then, the commutator of the hamiltonian with $\Phi$ is given by the commutator with $H^\partial$.  

On the other hand, we may consider the action of the hamiltonian on non gauge-invariant expressions, like $\phi(x)$ or $h_{\mu\nu}(x)$.  This may be understood by expanding in the metric perturbation,
\beq\label{Eexp}
G_{\mu\nu} + \Lambda g_{\mu\nu} = {\cal G}_{\mu\nu} - 8\pi G\, t_{\mu\nu}\ ,
\eeq
where ${\cal G}_{\mu\nu}$ is the linear term in $h$ and $t_{\mu\nu}$ contains quadratic and higher terms in $h$.  The first term, ${\cal G}_{\mu\nu}$, gives a total derivative which integrates to a surface term cancelling $H^\partial$,
\beq\label{derivreln}
\frac{1}{8\pi G} \int_\Sigma dV n^\mu   {\cal G}_{\mu\nu}\xi^\nu = H^\partial\ ,
\eeq
as may be seen by trivial generalization of the argument leading to (B24) of \cite{DoGi2}.  Then $H$ becomes
\beq\label{Hbulk}
H= \int_\Sigma dV n^\mu \left( T_{\mu\nu}+t_{\mu\nu}  \right)\xi^\nu\ ,
\eeq
and so is given in terms of the stress tensor $T_{\mu\nu}$ for matter and $t_{\mu\nu}$ for gravity.  Commutators of this expression with $\phi(x)$ or $h_{\mu\nu}(x)$ then generate their time translations via the canonical commutators.

In short, the boundary hamiltonian $H^\partial$ only correctly time translates gauge-invariant objects like $\Phi(x)$, and likewise the boundary stress tensor ${\cal T}_{\alpha\beta}$ is only expected to act correctly on such objects.  On the other hand, the bulk expression \eqref{Hbulk} correctly translates either gauge-invariant or gauge-variant objects; on the former the stress tensor acts to correctly translate the different pieces of $\Phi$.  

We can explicitly check to see that $H^\partial$ correctly time translates $\Phi$.  
To do so, it  needs to have the commutator
\beq\label{hvcomm}
[H^\partial,V^\tau(x)]= -i
\eeq
with the dressing.  This can be checked in the simple case of the line dressing \eqref{tcomp}.  To see this, note that the non-trivial part of the commutator comes from the time derivative of the metric perturbation in \eqref{tcomp} and the diagonal terms in \eqref{Hbdy}.  In particular, we find that
\beq\label{rrcom}
[h_{\rho\rho}(x),V_L^\tau(x')] = - \frac{i\kappa(D-3)}{2(D-2)R^{D-4}} \frac{\delta^{D-2}(\hat e - \hat e')}{\sqrt{\hat g}} \theta(\rho-\rho')(\sin\rho -\sin\rho') \frac{\cos^{D-5}\rho}{\sin^{D-2}\rho}
\eeq
and
\beq\label{trcom}
[{\hat g}^{ab}h_{ab}(x),V_L^\tau(x')] = \frac{i\kappa}{2R^{D-4}} \frac{\delta^{D-2}(\hat e - \hat e')}{\sqrt{\hat g}}\theta(\rho-\rho')(\sin\rho -\sin\rho') \frac{\cos^{D-5}\rho}{\sin^{D-4}\rho}\ .
\eeq
Using these to evaluate $H_\partial$ from \eqref{Hbdy} then yields \eqref{hvcomm}.  Notice also that the expressions \eqref{rrcom}, \eqref{trcom} are  useful in calculating the dressing field for an operator not located at the origin, generalizing the derivation of section \ref{Dressf}.

\section{Puzzles regarding gravitational holography and the boundary algebra}

It has been argued by Marolf\cite{MaroUH,Marothought,Maro2013} that the {\it origin} of holography, and thus of a proposed AdS/CFT equivalence, is intrinsically gravitational in nature, and in particular that it arises from the gauge symmetry and constraints of gravity, and {\it not} from stringy behavior such as extendedness of strings.  Since the preceding discussion describes how to solve the constraints to find gauge-invariant operators, in an order-by-order expansion in $\kappa$, and how to find the boundary symmetry generators, it provides a very concrete approach to assessing such statements.

\paragraph*{\underline{Bulk algebra, induced boundary algebra, and ``holography of information''}}

A starting point for the argument of  \cite{MaroUH,Marothought,Maro2013} is the relationship between the bulk and boundary algebras.  The bulk algebra is simply that of bulk operators.  In the case where we consider bulk operators that are not gauge-invariant and depend on the metric, defining the algebra may depend on the gauge-fixing prescription, as is for example described in appendix \ref{CQAdS}.  On the other hand, gauge-invariant operators should have no such dependence.\footnote{A possible alternative approach (see \cite{DoGi1}) is to work in a particular gauge, such as Fefferman-Graham gauge, and define the algebra via the Dirac brackets.}  Next, if boundary operators are obtained from the boundary limit of bulk operators, the bulk algebra {\it induces} an algebra of these boundary operators.  Of course the consistency and closure of this algebra depends on details of this boundary limit.  As an example, consider the operators ${\cal T}_{\alpha\beta}$, which were derived for $\alpha\beta= \tau\tau,\tau a$ in the preceding section.  These are interpreted as components of the boundary stress tensor, and have an algebra induced from that of the bulk perturbation $h_{\mu\nu}$.  The statement that the various generators $H_\xi$ of the $SO(D-1,2)$ symmetries should have the correct algebra presumably implies that the components of ${\cal T}_{\alpha\beta}$ have induced commutators corresponding to those of the stress tensor\footnote{For related discussion, see \cite{BoDe}.} of a conformal theory, although we have not checked this explicitly.

We next summarize  the basic argument for ``holography of information'' of \cite{Maro2013}.  Consider a bulk operator, such as $\Phi(\tau,x^i)$ of \eqref{giop}.  Suppose that this operator can be written in terms of operators in the boundary algebra, at some later time (or times) $\tau'$.   Then, since the boundary Hamiltonian $H^\partial$ of the preceding section is also in the boundary algebra, the resulting expression may be converted into an expression in terms of operators at time $\tau$.  This is done by conjugating by $\exp\{iH^\partial(\tau-\tau')\}$, which translates the operators at $\tau'$ to operators at $\tau$.  If all this can be done, then the operator $\Phi(\tau,x^i)$ has been rewritten in terms of boundary operators at the equal time $\tau$.  This, then, would explicitly exhibit the equal time ``holographic map'' between bulk and boundary operators, and so such an explicit expression would clearly be of significant interest.

The intuition that $\Phi(\tau,x^i)$ can be written in terms of boundary operators at a later time $\tau'$ arises from the intuition that this operator creates a state, which then propagates out to the boundary, and can be though of as arising from a boundary operator.  This kind of construction certainly holds in the free ($\kappa=0$) theory and has been explored in work going back to \cite{BDHM,HKLL,HMPS,KLgauge,KLgrav}; we refer to it as the HKLL construction.  A question is how such a construction would work in the interacting theory, since gravitational interactions also imply the presence of gravitational dressing and the need to properly handle the gauge structure of gravity -- as well as other subtleties of strong gravity.

\paragraph*{\underline{The extrapolate map?}}

A starting point for the HKLL construction is the ``extrapolate'' map, which in the free theory states how the field operator  $\phi(x)$ maps to a boundary operator in the limit as it approaches the boundary,
\beq\label{fext}
 {\cal O}(x^\alpha)=\lim_{\rho\rightarrow\pi/2} (\cos \rho)^{-\Delta} \phi(x^\alpha,\rho) \ ,
\eeq
where the conformal dimension $\Delta=(D-1+\sqrt{(D-1)^2+4m^2R^2})/2$ is given in terms of the mass $m$.

The analogous statement  of the extrapolate map in the $\kappa\neq0$ theory is plausibly also a good starting point there, but raises the question of the proper form of this map.  There are different possibilities.  

A first possibility is that one continues to use the map \eqref{fext}, in terms of the undressed operator.  However, if this were the case, then the boundary stress tensor \eqref{bdyT} and hamiltonian \eqref{Hbdy} would {\it commute} with ${\cal O}(x^\alpha)$ and wouldn't generate time translations of ${\cal O}(x^\alpha)$, as is seen explicitly in the discussion of the preceding subsection \ref{TBO}.

An alternate proposal is that the extrapolate map is formulated in terms of a dressed operator \eqref{giop},
\beq\label{gilim}
{\cal O}(x^\alpha)=\lim_{\rho\rightarrow\pi/2} (\cos \rho)^{-\Delta} \Phi(x^\alpha,\rho)\ .
\eeq
However, this raises new questions.  First, we have found {\it different} dressed operators -- {\it e.g.} with the line dressing or Coulomb dressing -- and there are  many more.  These appear to give different prescriptions.\footnote{An open question for the future is whether the boundary limit in the end suppresses   differences between these operators.}  Moreover, if we consider a particular form of the dressing, say the line dressing, then that linear structure is not maintained under time evolution, as was noted in section \ref{Dressf} -- a generic such dressing evolves by spreading out, approaching the Coulomb dressing. One could posit \eqref{gilim} using the Coulomb operator $\Phi_C$.  However, this operator has nonvanishing equal-time commutator with ${\cal T}_{\alpha\beta}(x^\gamma)$ over the entire boundary, and does not obviously localize in the boundary limit.

Indeed, a related question involves the locality of the operators.  Dressed operators do not commute outside the lightcone\cite{DoGi1} and so, as discussed above, their algebra is nonlocal.  This raises the question of locality of the corresponding boundary operators, which at $\kappa=0$ were argued to obey a local algebra in \cite{FKM}.  For example, in principle it looks like it may be possible to take a line operator $\Phi_L$ to an equal-time boundary point that is different from the point where the dressing is anchored; this could even be the antipodal point.  Such an operator would not commute with an operator at the spacelike-separated anchor point.

In short, the presence of gravitational dressing raises questions about how to properly and unambiguously define an extrapolate map that is valid to order $\kappa$ and beyond.  The identification \eqref{fext} leads to the wrong commutators, and \eqref{gilim} apparently suffers from ambiguities.

A related perspective on these puzzles it to try to understand how the boundary algebra and expression \eqref{fext} arise from the $\kappa\rightarrow0$ limit.\footnote{Note that since $\kappa$ is dimensionfull, it does not by itself provide a good expansion parameter.  One natural bulk expansion parameter is $\kappa  E^{(D-2)/2}$, where $E={\cal E}/R$ is the bulk energy conjugate to $t=R\tau$.  The corresponding CFT expansion parameter in the case of $AdS_5\times S^5$ is ${\cal E}^4/N$.} We find the correct commutator $[{\cal T}_{\alpha\beta},\Phi]$ by including the dressing, which is of order $\kappa^1$, in $\Phi$.  The limit produces a non-vanishing result for this commutator because the boundary stress tensor scales as $\kappa^{-1}$ (see \eqref{bdyT}, \eqref{Ttf}) -- and so this tensor lacks a clear definition in this limit.  This also raises the question of how to define the boundary algebra\cite{MaroUH} in this limit.

\paragraph*{\underline{The HKLL map?}}

Questions continue when one attempts to infer the $\calo(\kappa)$ and higher generalization of the HKLL map, which we have seen plays a key role in the boundary unitarity argument of \cite{MaroUH,Marothought,Maro2013}.  First, given the apparent ambiguity in the extrapolate map, it is also not clear how to invert it to provide a construction of $\Phi(\tau,x^i)$ in terms of operators $\calo(x^\alpha)$ at a given time.  In principle, it has been advocated that this construction arises via evolution by the bulk equations of motion, which in the free case propagate the particle to the boundary.  How this works is less clear working to order $\kappa$ and beyond.  For example, if we consider the line operator $\Phi_L$ creating a particle at $\vec x\neq 0$, we have seen that this operator can be interpreted as creating the particle with the Coulomb dressing, together with  a radiative component of the gravitational field.  The corresponding gravitational radiation does not all necessarily reach the boundary at the same time that the particle would.  In order to find a boundary dual of each of the different bulk operators $\Phi$, one 
one could try to find an order-by-order relation of perturbative bulk fields to boundary operators, by solving bulk equations of motion, as  in \cite{KLgauge,KLgrav,HMPS}. An open question is whether this provides a systematically well-defined prescription including dressing effects.\footnote{Note also the question of how to treat the metric perturbations, given that they fall off more slowly than the typically assumed behavior \eqref{HeTeBC}.}  

Of course, part of the challenge is determining the evolution of a dressed operator $\Phi(x)$.  In principle, its Heisenberg equation of motion can be written as 
\beq
{\ddot \Phi} = - [H,[H,\Phi]]\ .
\eeq
and this has formal solution
\beq
\Phi(\tau)= e^{i\tau H^\partial } \Phi(0) e^{-i\tau H^\partial}\ ,
\eeq
but finding an explicit solution is a challenge.
Alternately, one can try to read off the equations for $\Phi$ from its expression in terms of $\phi(x)$ and $h_{\mu\nu}(x)$, and from the bulk equations for these.  But, in either case the equations that result are apparently not simple, and require solving the constraints. 

These points raise important questions about the precise form of a generalization of the HKLL map to $\calo(\kappa)$ and beyond\footnote{Note that aspects of such a map have been proposed in \cite{KLgrav}.  This work was based in part on the assumption of commutativity of the bulk observables, which contrasts with the non-commutative algebraic structure that we have found for the bulk observables.} and reinforce the apparent necessity of solving the full bulk equations.  Related discussion of questions in time-dependent backgrounds appears in \cite{AAL}.

\paragraph*{\underline{Boundary unitarity and holography of information?}}

Without a complete prescription for writing $\Phi(\tau,x^i)$ in terms of operators $\calo(x^\alpha)$ at a given ({\it e.g.} later) time, it is not clear how to sharply formulate the argument for holography of information; more definite expressions, valid to $\calo(\kappa)$ and beyond, appear to be needed to assess its viability.  And if, for example, an order-by-order procedure as noted above is carried out, using HKLL inversion of expressions such as \eqref{fext}, then it is unclear how to use the step in the holography argument that relies on $H^\partial$ generating time translations, since that is only true for the action on gauge-invariant operators, and not on perturbative fields as in \eqref{fext}.  Perturbative fields are instead translated by the full bulk hamiltonian, \eqref{Hbulk}.  Put differently, the Hamiltonian $H$ can only be thought of as lying in the boundary algebra when acting on operators commuting with the constraints, and this amounts to solving the equations for unitary time evolution in the bulk.

Of course, another closely related possible approach to achieving the goal of these arguments -- expressing $\Phi(\tau,x^i)$ in terms of operators $\calo(x^\alpha)$ {\it at equal time} -- is to consider conjugating $\Phi(x)$ by the translation operator; in the limiting case, such a large translation reaches the boundary, and thus 
could directly relate $\Phi(x)$ to a boundary operator.  

However, as was argued in \cite{DoGi3}, performing such a large translation again requires having the dressed operator $\Phi(x)$ to all orders in $\kappa$.  This would again be equivalent to solving the full nonperturbative generalization of the constraint equations $G_{0\mu}+\Lambda g_{0\mu}=8\pi G T_{0\mu}$, and this is equivalent to having the full non-perturbative evolution of the system, since solving the constraints amounts to solving for time evolution.\footnote{In the context of a hypothesized more complete quantum gravity theory, the constraints \eqref{Constraints} may well be replaced by some more complete evolution equation and/or expression of gauge invariance.  For one example of a possible role for high-energy degrees of freedom, see \cite{Harl}.}

Indeed, an important claim of the AdS/CFT correspondence -- if it truly defines a fine-grained quantum {\it equivalence} of theories\cite{AdSqs} -- is that it maps boundary evolution, which is manifestly unitary, onto bulk evolution, and thus it would appear to demonstrate the form of unitary bulk evolution.  This is particularly important in view of the unitarity crisis (information problem) for black holes, which raises the profound question of how black hole evolution {\it can} be unitary.  Specifically, if we were given the precise map between bulk and boundary theories, this should map unitary boundary evolution onto unitary bulk evolution.  But, the preceding arguments suggest that the problem of {\it defining} the ``holographic map'' is not independent of the problem of describing unitary evolution in the bulk -- finding this map and finding the form of unitary bulk evolution are directly linked.  In this sense such an AdS/CFT correspondence doesn't obviously solve the problem of bulk unitarity and in particular the unitarity crisis for black holes -- it requires its solution.

\begin{acknowledgments}
We would like to thank W. Donnelly, D. Harlow, G. Horowitz, D. Marolf, and S. Weinberg for useful conversations.  
This material is based upon work supported in part by the U.S. Department of Energy, Office of Science, under Award Number {DE-SC}0011702, by Foundational Questions Institute (fqxi.org) Grant No. FQXi-RFP-1507, and by the National Science Foundation Graduate Research Fellowship Program under Grant No. 1650114.

\end{acknowledgments}

\appendix

\section{Canonical quantization in AdS}
\label{CQAdS}

To quantize metric fluctuations in AdS, a first step is to find the canonical commutators arising from the action \eqref{gravact}, again working perturbatively in the expansion \eqref{metpert}.  This is most easily done by writing the metric \eqref{metpert} in Arnowitt-Deser-Misner (ADM) form\cite{ADM},
\beq
ds^2= -N^2d\tau^2 + q_{ij}(dx^i + N^i d\tau)  (dx^j + N^j d\tau)\ .
\eeq
One easily finds from \eqref{metpert} the relation to the metric perturbation,
\beq
q_{ij}=g^0_{ij}+\kappa h_{ij}\quad ,\quad N^i = \kappa q^{ij}h_{j\tau} \quad ,\quad N^2 = -g^0_{\tau\tau} - \kappa h_{\tau\tau} + \kappa^2 q^{ij} h_{i\tau}h_{j\tau}\ .
\eeq
The spatial metric $q_{ij}$ has conjugate momenta
\beq\label{canonp}
\pi^{ij} = \frac{2}{\kappa^2} \sqrt{q} (K^{ij}-q^{ij} K) + \frac{1}{\alpha\kappa^2}\frac{\sqrt{|g^0|}}{|g|}n_\nu \nabla^0_\mu\left(\sqrt{|g|}g^{\mu\nu}\right)\sqrt q q^{ij}\ ,
\eeq
with extrinsic curvature defined as
\beq
K_{ij}= \frac{1}{2N} (\dot q_{ij} - D_iN_j- D_j N_i)\ ,
\eeq
covariant derivative $D_i$ defined with respect to the spatial metric $q_{ij}$, and normal $n^\mu=(1/N,-N^i/N)$ to the constant-$\tau$ slices.  
Eq.~\eqref{canonp} agrees with the usual ADM momentum for vanishing gauge-fixing term ($\alpha=\infty$).  The extra term is proportional to the gauge term 
$\nabla^0_\mu\left(\sqrt{|g|} g^{\mu\nu}\right)$; this generates gauge transformations and so may be dropped in commutators with gauge-invariant objects; alternately, it vanishes in the gauge symmetry restoring limit $\alpha\rightarrow\infty$.
If the action is written in ADM form the momenta conjugate to $N,N^i$ vanish, up to such a gauge-generating term proportional to $1/\alpha$.

The canonical commutators  take the form
\bea
\left[q_{ij}(\tau,\vec{x}),q_{kl}(\tau,\vec{x}')\right] & = & 0\\
\left[\pi^{ij}(\tau,\vec{x}),\pi^{kl}(\tau,\vec{x}')\right] & = & 0\\
\left[q_{ij}(\tau,\vec{x}),\pi^{kl}(\tau,\vec{x}')\right] & = & i\delta_{i}^{(k}\delta_{j}^{l)}\delta^{D-1}(\vec{x}-\vec{x}') \label{pqc}
\eea
where $\delta^{D-1}$ is the Dirac delta function and our symmetrization normalization is $(ij)=(ij+ji)/2$.  In the presence of the $\alpha$ term, \eqref{pqc} may be extended to 
\beq
\left[g_{\mu\nu}(\tau,\vec{x}),\pi^{\lambda\sigma}(\tau,\vec{x}')\right] = i\delta_{\mu}^{(\lambda}\delta_{\nu}^{\sigma)}\delta^{D-1}(\vec{x}-\vec{x}')
\eeq

Written in terms of the metric perturbation, the  momenta \eqref{canonp} with $\alpha=\infty$ become 
\beq
\pi^{ij}=\frac{1}{\kappa N} \sqrt{q} q^{ik}q^{jl} \left[\dot h_{kl} - D_k h_{l\tau} -D_l h_{k\tau} - q_{kl} q^{mn}(\dot h_{mn} -2D_m h_{n\tau} ) \right]\ .  
\eeq
Thus the metric perturbations commute at equal times, and satisfy
\beq
\left[h_{ij}(\tau,\vec{x}), \dot h_{kl}(\tau,\vec{x}')-q_{kl} q^{mn}\dot h_{mn}(\tau,\vec{x}')\right]= iN \frac{(q_{ik}q_{jl}+q_{il}q_{jk})}{2}  \frac{\delta^{D-1}(\vec{x}-\vec{x}')}{\sqrt q}\ ,
\eeq
 or
\beq\label{hhcomm}
\left[h_{ij}(\tau,\vec{x}), \dot h_{kl}(\tau,\vec{x}')\right]= iN \frac{\delta^{D-1}(\vec{x}-\vec{x}')}{\sqrt q}\left( \frac{q_{ik}q_{jl}+q_{il}q_{jk}}{2} -\frac{q_{ij}q_{kl}}{D-2}\right)\ .
\eeq

\section{Dressing theorem in AdS}

This appendix generalizes the {\it dressing theorem} of \cite{DoGi2} to the case of  asymptotically AdS spacetimes.  
 Let ${\cal O}$ be a compactly supported operator in
a quantum field theory in an AdS background. Let $\tilde{{\cal O}}$
be a gravitationally dressed (i.e. gauge-invariant) version of ${\cal O}$,
which has perturbative expansion
\beq
\tilde{{\cal O}}={\cal O}+\kappa{\cal O}^{(1)}+\kappa^{2}{\cal O}^{(2)}+\cdots
\eeq
Then the theorem states that ${\cal O}$ transforms nontrivially under some AdS isometry if
and only if ${\cal O}^{(1)}$ depends on the asymptotic spacetime
metric. 

To prove this, note that gauge invariance states that $\tilde{\cal O}$ commutes with the constraints,
\beq
\left[\int_{\Sigma}C_{\xi},\tilde{{\cal O}}\right]=0\ ,
\eeq
with $C_\xi$ given in \eqref{Ceq}, for any compactly supported $\xi$; this must therefore also hold for $\xi$ with non-compact support.
We consider this statement for an infinitesimal AdS isometry $\xi$. 
Once again, we expand the Einstein tensor as in \eqref{Eexp}, so that $C_\xi$ becomes
\beq
C_{\xi}=(-1)^{D}\star\left[\left(-\frac{1}{8\pi G}{\cal G}_{\mu\nu}+T_{\mu\nu}+t_{\mu\nu}\right)\xi^{\nu}dx^{\mu}\right]\ ,
\eeq
with $t_{\mu\nu}$ the  gravitational stress tensor defined in \eqref{Eexp}. 
Next, integrating over the surface $\Sigma$ gives
\beq
\int_{\Sigma}C_{\xi}=-H^\partial_{\xi}+\int_{\Sigma}dV \left(T_{\mu\nu}+t_{\mu\nu}\right)n^{\mu}\xi^{\nu}\ .
\eeq
Here we have used the generalization of \eqref{derivreln}
\beq
H^\partial_\xi=\frac{1}{8\pi G} \int_\Sigma dV n^\mu {\cal G}_{\mu\nu} \xi^\nu\ ,
\eeq
which is true for a Killing vector $\xi^\mu$; this is shown via a simple generalization of the proof of the analogous flat space relation, (B25), given in \cite{DoGi2}.

The statement of gauge invariance then becomes
\beq
\left[-H^\partial_{\xi}+\int_{\Sigma}dV\left(T_{\mu\nu}+t_{\mu\nu}\right)n^{\mu}\xi^{\nu},\tilde{{\cal O}}\right]=0\ .
\eeq
This statement can be expanded
\beq
\left[H^\partial_{\xi},\kappa{\cal O}^{(1)}+\kappa^{2}{\cal O}^{(2)}+\cdots \right]=\left[\int_{\Sigma}dV\left(T_{\mu\nu}+t_{\mu\nu}\right)n^{\mu}\xi^{\nu},{\cal O}+ \kappa{\cal O}^{(1)}+\kappa^{2}{\cal O}^{(2)}+\cdots\right]\ ,
\eeq
where, for a localized operator $\cal O$, $[H_\xi^\partial,\calo]=0$.  Since $H^\partial_{\xi}$ is $\calo(1/\kappa)$ if $\calo$ is independent of $h_{\mu\nu}$, the $\calo(\kappa^0)$ term in this equation states
\beq
\left[H^\partial_{\xi},\kappa\tilde{{\cal O}}^{(1)}\right]=\left[\int_{\Sigma}dV T_{\mu\nu}n^{\mu}\xi^{\nu},{\cal O}\right]
\eeq
The commutator on the left vanishes if and only if the commutator on the right does.  
But since $H^\partial_{\xi}$ depends only on the boundary metric,
the commutator on the left is nonvanishing if and only if $\tilde{{\cal O}}^{(1)}$
depends on the asymptotic metric. The commutator on the right is just
that of the $\calo(\kappa^0)$  AdS isometry generator with ${\cal O}$, which vanishes
if and only if ${\cal O}$ has zero charges under the AdS isometry
group. This proves the theorem. 

\section{Boundary conditions for metric perturbations}\label{Falloffs}

The main text found a  difference between the falloff found for the dressing field, \eqref{dressfieldr}, \eqref{dressfielda}, and the boundary conditions for a ``standard'' form of AdS/Schwarzschild, \eqref{Schpert}, which were advocated as more general boundary conditions in \cite{HeTe}; there is a    related question of defining charges.   This appendix briefly describes some aspects of boundary conditions.

Specifically, two important criteria for boundary conditions of a self-contained system is that they give finite symplectic form, and vanishing asymptotic symplectic flux so that the symplectic form is conserved. The symplectic form is
\beq
\Omega=\int_\Sigma \delta\theta\ ,
\eeq
with $\Sigma$ a spatial slice and symplectic potential $\theta$ given in \eqref{sympot}.  Moreover, if $S_\rho$ is a constant $\rho$ surface that approaches the boundary of AdS as $\rho\rightarrow\pi/2$, the condition of vanishing asymptotic flux is
\beq
\lim_{\rho\rightarrow\pi/2} \int_{S_\rho} \delta\theta =0\ .
\eeq
An explicit formula for the integrands of these is\cite{Crnkovic:1986ex,BuWa}
\beq
\star\delta\theta^\iota(\delta g^1,\delta g^2)= \frac{2}{\kappa^2} P^{\iota\lambda\mu\nu\sigma\upsilon}\left(\delta g^2_{\lambda\mu}\nabla_\nu \delta g^1_{\sigma\upsilon}-\delta g^1_{\lambda\mu}\nabla_\nu \delta g^2_{\sigma\upsilon}\right)\ ,
\eeq
with
\beq
P^{\iota \lambda \mu \nu \sigma \upsilon}=g^{\iota \sigma }g^{\upsilon\lambda }g^{\mu \nu}-\frac{1}{2}g^{\iota \nu}g^{\lambda \sigma }g^{\upsilon\mu }-\frac{1}{2}g^{\iota \lambda }g^{\mu \nu}g^{\sigma \upsilon}-\frac{1}{2}g^{\lambda \mu }g^{\iota \sigma }g^{\upsilon\nu}+\frac{1}{2}g^{\lambda \mu }g^{\iota \nu}g^{\sigma \upsilon}\ .
\eeq

Starting with these expressions and the form \eqref{adsmet} of the AdS metric, carefully enumeration of the different terms shows sufficient conditions for finiteness of $\Omega$ and vanishing of the flux.  In particular, this shows that the behavior $h_{\rho\rho}\sim h_{ab}\sim \cos^{D-5}\rho$, as found in \eqref{dressfieldr}, \eqref{dressfielda}, yield finite symplectic form for $D>3$.  The conditions for vanishing asymptotic flux can also be checked; in this paper we take $D\geq4$ to ensure finiteness and conservation.

\bibliography{dress}

\end{document}